\documentclass[hyper]{JHEP3} % 10pt is ignored!

\usepackage{epsfig}

\newcommand\fverb{\setbox\pippobox=\hbox\bgroup\verb}

\newcommand\fverbdo{\egroup\medskip\noindent%

            \fbox{\unhbox\pippobox}\ }

\newcommand\fverbit{\egroup\item[\fbox{\unhbox\pippobox}]}

\newbox\pippobox

\title{Current Algebra and Integrability
 of Principal
Chiral Model on the World-sheet with 
General Metric}

\author{J. Kluso\v{n}
 \footnote{On leave from Masaryk University, Brno}\\
Dipartimento di Fisica \& Sezione I.N.F.N.\\
Universit\`a di Roma
``Tor Vergata'' \\
Via della Ricerca Scientifica 1 00133  Roma   ITALY\\
E-mail:
\email{Josef.Kluson@roma2.infn.it}}
\preprint{
\hepth{0703003}}
\abstract{We  study the classical current
algebra  for principal
chiral model defined on two dimensional world-sheet with general
 metric. We develop the Hamiltonian formalism and
determine the form of the Poisson brackets between currents.
Then we determine the Poisson bracket for Lax connection and
we  show that this Possion bracket does not depend on the
world-sheet metric. 
We  also study the Nambu-Gotto
form of this model. We  prove an existence of the Lax
connection and  determine their Poisson bracket.}


\newcommand{\mT}{\mathcal{T}}
\def\tr{\mathrm{Tr}}

\def\pb  #1{\left\{#1\right\}}

\newcommand{\bA}{{\bf A}}

\newcommand{\mH}{\mathcal{H}}

\newcommand{\mL}{\mathcal{L}}

\newcommand{\hJ}{\hat{J}}

\newcommand{\bAi}{\left(\bA^{-1}\right)}

\newcommand{\tgamma}{\tilde{\gamma}}
\begin{document}
%%%%%%%%%%%%%%%%%%%%%
%%%%Introduction %%%%%%%%%
%%%%%%%%%%%%%%%%%%%%
\section{Introduction}
The discovery of the integrability of
the classical string sigma model on
$AdS_5\times S^5$ was one of the great
achievements in the last few years
\cite{Bena:2003wd}
\footnote{For some works considering
integrability of sigma model on $AdS_5\times
S^5$, see
\cite{Kluson:2007gp,Grassi:2006tj,
Dorey:2006mx,Gromov:2006dh,Alday:2005ww,Das:2005hp,Alday:2005jm,
Arutyunov:2005nk,Frolov:2005dj,Chen:2005uj,Alday:2005gi,Das:2004hy,
Arutyunov:2004yx,Beisert:2004ag,Kazakov:2004nh,Berkovits:2004jw,
Hatsuda:2004it,Kazakov:2004qf,Alday:2003zb,Vallilo:2003nx}}.
More precisely, the authors  
\cite{Bena:2003wd} found a Lax formulation of the equations
of motion that leads to the existence of 
an infinite tower of conserved
charges in the classical world-sheet theory. 
Using these charges it was possible to construct
and classify large families of exact 
solutions of the classical equations of motion. 
However as was recently stressed in a very nice paper
\cite{Dorey:2006mx} this fact does not 
quite coincide with the standard 
definition of integrability. Integrability 
in the standard sense requires not only the existence
of a tower of conserved charges 
but also requires that these charges
be in involution. In other words 
the conserved charges should Poisson 
commute with each other.  The knowledge
of the Poisson brackets is necessary 
for constructing the action-angle variables
for the system that plays an important
role in semi-classical quantisation. 
The careful analysis presented 
in \cite{Dorey:2006mx} explicitly 
demonstrated that for classical string
moving on $R\times S^3$ submanifold 
of $AdS_5\times S^5$ that the Poisson
brackets of conserved charges are in involution.

However it should be stressed that even
if the bosonic string on $AdS_5\times S^5$
admits Lax formulation, there is a well
known problem in determining the Poisson
brackets of conserved charges. The problem
is due to the presence of Non-Ultra 
Local terms in the Poisson brackets of the
world-sheet fields that lead to the
ambiguities in brackets for the charges. 
Some proposals  for resolution
of this issue  were recently discussed in
\cite{Das:2004hy,Dorey:2006mx}. 
For example, it was shown 
in \cite{Dorey:2006mx} that there exists
the resolution of this problem 
 using the prescription of for
 regularising of the 
the problematic brackets that 
was introduced in the
 earlier work by Maillet
\cite{Maillet:1985ek,Maillet:1985ec}.
Explicitly, it was
demonstrated in \cite{Dorey:2006mx}
that  the application
of this procedure for the bosonic
string in the conformal gauge moving
on an $R\times S^3$ submanifold of the
full $AdS_5\times S^5$
 leads to the natural
symplectic structure on the space
of finite-gap solutions of the
equations of motion that were constructed in
\cite{Dorey:2006zj}. 

As the next step 
it is natural to ask the question whether the
 principal chiral
model is integrable on the
 world-sheet with general metric.
 An importance of the
fact that one can define the Lax connection
on the world-sheet with general metric
 was stressed in \cite{Arutyunov:2004yx}.
 On the other hand the form of the
current algebra for general metric is
not known. Consequently it is  also not known
 the form of the Poisson
bracket of the Lax connection and hence
it is not completely clear whether one
can define infinite number of conserved
charges that are in involution on the
general world-sheet as well. 

The goal of this paper is to 
answer these questions. 
We explicitly calculate the Poisson
bracket of the principal chiral model
on the world-sheet with general two
dimensional metric. Even if the
resulting Poisson bracket are complicated
we will see that the Poisson bracket
of the Lax connection does not depend
on the metric at all and takes precisely
the same form as was recently reviewed
in \cite{Dorey:2006mx}. According
to the arguments given there it follows
that the principal chiral model defined
on the world-sheet with the general
metric possesses again infinite number
or conserved charges that are in involution.

It is important to stress that
this result holds for the principal chiral
model where the  gauge symmetries
are not  fixed at all.
The situation becomes  more involved
in case when the gauge fixing functions depend on 
the phase space variables. An example of
 such  a more complicated gauge
is  \emph{uniform light cone}
\cite{Arutyunov:2006gs,Arutyunov:2005hd}
\footnote{For recent discussion of this
gauge, see for example
\cite{Klose:2006zd,Astolfi:2007uz}.}.
The consistency of this gauge fixing implies
that the Lagrange multipliers are completely
fixed, the Virasoro constraints together
with the gauge fixing conditions become
the second class constraints and finally
the original Hamiltonian strongly vanishes.
In order to appropriate study this theory
we should identify the Hamiltonian on the
reduced phase space and also determine
the Dirac brackets for the phase space
variables. We leave the study of integrability
and the current algebra for this gauge
fixing theories for future.

We  also discuss the Nambu-Gotto form
of the principal chiral model. We 
demonstrate that it is again integrable
and we will calculate the algebra of 
the currents. We also calculate the
Poisson bracket of the Lax connection
and we show that it takes the same
form as in the paper 
\cite{Dorey:2006mx}. This result again
demonstrates an integrability of this
form of the principal chiral model. 

The organisation of the paper is as follows.
In the next section (\ref{second})
we introduce the basic notation of the
principal chiral model defined on 
two dimensional world-sheet with general
metric. In section (\ref{third}) we 
calculate the algebra of currents. We 
also calculate the Poisson bracket of 
spatial component of the Lax connection.
In section (\ref{fourth}) we 
perform the Hamiltonian analysis of
the Nambu-Gotto form of the principal
chiral model. We  show that this
model is again integrable and we determine
the algebra of current and also the 
Poisson bracket of the Lax connection
for this model. In conclusion
(\ref{fifth}) we outline our results
and suggest further directions of
research.
%%%%%%%%%%%%%%%%%%%%%%%%%%%%%%%%%%
%%%%%%%%%%%%%%%%%%%%%%%%%%%%%%%%%%%%
\section{Hamiltonian formalism
for principal chiral model}\label{second}
In this section we introduce the
Hamiltonian formalism for the principal
chiral model defined on two-dimensional
world-sheet with general metric $\gamma$. 
The action for this model takes the form
\begin{eqnarray}\label{Sact}
S&=&-\frac{\sqrt{\lambda}}{4\pi}
\int d\tau d\sigma \sqrt{-\gamma}
\gamma^{\alpha\beta} J^A_\alpha J_\beta^BK_{AB}=
\nonumber \\
&=&
-\frac{\sqrt{\lambda}}{4\pi}
\int d\tau d\sigma \sqrt{-\gamma}
\gamma^{\alpha\beta}
E^A_M\partial_\alpha x^M E_N^B\partial_\beta x^N
K_{AB}=\nonumber \\
&=&-\frac{\sqrt{\lambda}}{4\pi}
\int d\tau d\sigma \sqrt{-\gamma}
\gamma^{\alpha\beta}
G_{MN}\partial_\alpha x^M \partial_\beta x^N \ . 
\nonumber \\ 
 \end{eqnarray}
 Here $\frac{\sqrt{\lambda}}{2\pi}$ 
 is the effective string tension. Coordinates $
 \sigma$ and $\tau$ parametrise the string
 world-sheet.  We assume the range of
 $\sigma$ to be $-\pi \leq \sigma \leq \pi$.
The current $J_\alpha^A$ is defined through
the group element $g$ that belongs to 
the group $G$ as 
\begin{equation}\label{defJ}
J_\alpha=g^{-1}\partial_\alpha g
\equiv J^A_\alpha T_A  \ , 
\end{equation} 
where $T_A$ form the basis of the algebra
$\mathbf{g}$ that obey the relations
\begin{equation}
\tr (T_A T_B)=K_{AB} \ , 
\quad [T_A,T_B]=f_{AB}^C T_C \ , 
\end{equation}
where $K_{AB}$ is invertible matrix
and where $f_{AB}^C=-f_{BA}^C$ are
structure constants of the algebra $\mathbf{g}$. 
 The
indices $A,B$ label components of the basis
$T_A$. If we parametrise the group element
with the fields $x^M$ we can write the
current $J_\alpha^A$ as
\begin{equation}
J_\alpha^A=
E^A_M\partial_\alpha x^M  \ . 
\end{equation}
Finally we have introduced the metric 
\begin{equation}
G_{MN}=
E_M^A K_{AB}E^B_N \ 
\end{equation}
defined on some target manifold
labelled with coordinates $x^M$. In this
interpretation $E_M^A$ are vielbeins of the
target manifold
\cite{Evans:1999mj}.

We now develop the Hamiltonian formalism for 
this principal chiral model, where we consider
$x^M$ and corresponding momenta $p_M$ as
canonical variables. There also  exists an
alternative form of the Hamiltonian formalism
based on presumption that the current $J_\sigma^A$
is the dynamical variable 
\cite{Faddeev:1987ph} (For some recent works,
see for example 
\cite{Das:2004hy,Korotkin:1997fi,Kluson:2007gp}).
It can be shown that these two approaches give
equivalent result. 
 
The conjugate momenta to $x^M$ are equal to
\begin{equation}
p_M=-\frac{\sqrt{\lambda}}{2\pi}
\sqrt{-\gamma}\gamma^{\tau\alpha}
E_M^AK_{AB}J^B_\alpha 
\end{equation}
and together with $x^M$ obey the canonical
Poisson brackets
\begin{equation}
\pb{x^M(\sigma),p_N(\sigma')}=
\delta^M_N\delta(\sigma-\sigma') \ . 
\end{equation}
% Note that using the definition above
% we can temprarly write 
% \begin{equation}
% \partial_\tau x^M=
% -\frac{1}{\sqrt{-\gamma}\gamma^{\tau\tau}}
% \frac{2\pi}{\sqrt{\lambda}}G^{MN}p_N
% -\frac{\gamma^{\tau\sigma}}{\gamma^{\tau\tau}}
% \partial_\sigma x^M
% \end{equation}
Then the   Hamiltonian  takes the form
\begin{eqnarray}
H_0&=&\int_0^{2\pi}d\sigma \mH_0=\int_0^{2\pi}
d\sigma[\partial_\tau x^M p_M-\mL]=
\nonumber \\
&=&\int_0^{2\pi}
d\sigma\left[
-\frac{1}{\sqrt{-\gamma}\gamma^{\tau\tau}}
\left(\frac{\pi}{\sqrt{\lambda}}G^{MN}p_Mp_N+
\frac{\sqrt{\lambda}}{4\pi}G_{MN}\partial_\sigma
x^M\partial_\sigma x^N\right)-\frac{\gamma^{\tau\sigma}}
{\gamma^{\tau\tau}}p_M\partial_\sigma x^M\right] \ . 
\nonumber \\
\end{eqnarray}
Note that the action is invariant
under Weyl transformation 
\begin{equation}\label{Weyltr}
\gamma_{\alpha\beta}'(\sigma,\tau)=
e^{\phi(\sigma,\tau)}\gamma_{\alpha\beta}
(\sigma,\tau) \ . \quad
% \sqrt{-\gamma}'=e^\phi\sqrt{-\gamma} \ .
\end{equation}
In what follows we use following
parametrisation of the metric
variables $\gamma_{\alpha\beta}$
\begin{equation}
\lambda^\pm=\frac{\sqrt{-\gamma}
\pm \gamma_{\tau\sigma}}{\gamma_{\sigma\sigma}} \ , 
\quad \xi=\ln \gamma_{\sigma\sigma} \ ,
\end{equation}
where $\lambda^\pm$ are manifestly
invariant under (\ref{Weyltr}) while
$\xi$ transforms as $\xi'(\sigma,\tau)=
\xi(\sigma,\tau)+\phi(\sigma,\tau)$. 
Then the Hamiltonian
density $\mH_0$ is equal to
\begin{eqnarray}
\mH_0&=&\frac{\lambda^++\lambda^-}{2}
T_0+\frac{\lambda^+-\lambda^-}{2}T_1=
\lambda^+T_++\lambda^-T_- \ , \nonumber \\
T_+&=&\frac{1}{2}(T_0+T_1) \ , \quad 
T_-=\frac{1}{2}(T_0-T_1) \ ,  \nonumber \\
\end{eqnarray}
where
\begin{eqnarray}
T_1&\equiv& 
p_M\partial_\sigma x^M \ , 
\nonumber \\
T_0&\equiv& 
\frac{\pi}{\sqrt{\lambda}}p_MG^{MN}p_N
+\frac{\sqrt{\lambda}}{4\pi}
\partial_\sigma x^M G_{MN}
\partial_\sigma x^N \ . \nonumber \\
\end{eqnarray}
Since the action (\ref{Sact}) does
not contain time-derivative of $\gamma_{\alpha\beta}$
it follows that the momenta conjugate
to $\lambda^\pm,\xi$ are zero:
\begin{equation}
\pi^\pm=\frac{\delta S}{\delta 
\partial_\tau\lambda^\pm}=0 \ , \quad
\pi^\xi=\frac{\delta S}{\delta
\partial_\tau \xi}=0 \ . 
\end{equation}
These conditions are primary constraints
of the theory. 
According to the general theory of the
constraint systems 
\cite{Henneaux:1992ig,Govaerts:1991gd,Govaerts:2002fq}
the primary constraints should be preserved
during the time evolution of the
system 
\begin{equation}
\dot{\pi}_\pm=\pb{\pi_\pm,H_0}=
T_\pm\approx 0 \ . 
\end{equation}
Then the  consistency of the theory implies
 two secondary constraints
\begin{equation}
T_+=T_-=0 \ . 
\end{equation}
Now we have to calculate the
Poisson brackets between constraints
$T_1(\sigma),T_0(\sigma')$
 and also between
$p_N,x^M$ and $T_0,T_1$.
\begin{eqnarray}
\pb{p_N(\sigma),T_1(\sigma')}&=&
-p_N(\sigma')\partial_{\sigma'}
 \delta(\sigma-\sigma')
%\partial_y P_N(y)\delta(x-y) \ , 
\nonumber \\
\pb{p_N(\sigma),T_0(\sigma')}&=&
% \frac{1}{2\sqrt{-\gamma}}
-\frac{\pi}{\sqrt{\lambda}}
p_K(\sigma')\partial_NG^{KL}p_L(\sigma')
\delta(\sigma-\sigma')-
\nonumber \\
&-&
\frac{\sqrt{\lambda}}{2\pi}\partial_{\sigma'}
\delta(\sigma-\sigma')G_{NK}(\sigma')
\partial_{\sigma'} x^K(\sigma')-
\nonumber \\
&-&
\frac{\sqrt{\lambda}}{4\pi}\partial_{\sigma'}
x^K(\sigma') 
\partial_N G_{KL}\partial_{\sigma'} x^L(\sigma')
\delta(\sigma-\sigma') \ , \nonumber \\
 \pb{x^N(\sigma),T_1(\sigma')}&=&
%&\frac{1}{\gamma_{11}}
\partial_{\sigma'} x^N(\sigma')\delta(\sigma-\sigma') \ , 
\nonumber \\
\pb{x^N(\sigma),T_0(\sigma')}&=&
% \frac{1}{2\sqrt{-\gamma}}
\frac{2\pi}{\sqrt{\lambda}}
G^{NL}p_L(\sigma')\delta(\sigma-\sigma') \ . 
\nonumber \\
\end{eqnarray}
Then it is easy to determine Poisson
brackets between $T_1,T_0$
\begin{eqnarray}
\pb{T_1(\sigma),T_1(\sigma')}&=&
\partial_\sigma T_1(
\sigma)\delta(\sigma-\sigma')+
2T_1(\sigma)\partial_\sigma \delta(\sigma-\sigma')
 \ , 
\nonumber \\
\pb{T_1(\sigma),T_0(\sigma')}&=&
\partial_\sigma T_0(\sigma)
\delta(\sigma-\sigma')+
2T_0(\sigma) \partial_\sigma
\delta(\sigma-\sigma') \  ,
\nonumber \\
\pb{T_0(\sigma),T_0(\sigma')}&=&
\partial_\sigma T_1(\sigma)
\delta(\sigma-\sigma')+
2T_1(\sigma)\partial_\sigma
\delta(\sigma-\sigma')\ .  \nonumber \\
\end{eqnarray}
Using these Poisson brackets we finally
get 
\begin{eqnarray}
\pb{T_+(\sigma),T_+(\sigma')}&=&
\partial_\sigma T_+(\sigma)
\delta(\sigma-\sigma')+
2T_+(\sigma)\partial_\sigma\delta(\sigma-\sigma') \ , 
\nonumber \\
\pb{T_-(\sigma),T_-(\sigma')}&=&
-\partial_\sigma T_-(\sigma)
\delta(\sigma-\sigma')-2
T_-(\sigma)\partial_\sigma
\delta(\sigma-\sigma') \ , 
\nonumber \\
\pb{T_+(\sigma),T_-(\sigma')}&=&0 \ .
\nonumber \\
\end{eqnarray}
These results imply that  the constraints
$T_0,T_1$ do not generate any additional ones.
We denote all constraints
as $\Phi_\alpha$, 
where $\alpha=(\lambda^\pm,\xi,\pm)$ label the
first class constraints $\pi^\lambda_\pm=
\pi_\xi=0$ and $T_\pm=0$. We introduce
the Lagrange multipliers $\rho^\alpha$.
Then the  generalised
Hamiltonian density takes the form
\begin{equation}\label{mHT}
\mH_T=\mH_0+
\rho^\alpha(p,x)\Phi_\alpha  \ ,
\end{equation}
where in principle $
\rho^\alpha$ depend on the phase
space variables $p_M,x^M,\pi^\lambda_\pm,
\lambda^\pm,\pi_\xi,\xi$.
Then (\ref{mHT})
implies that the equation
of motion for any $F$ defined on the phase space
takes the form
\begin{eqnarray}
\frac{dF}{d\tau} &=&\partial_\tau F+
\pb{F,H_T}=\nonumber \\
&=&\partial_\tau F+
\pb{F,H_0}+\int d\sigma\left(
\rho^\alpha (\sigma) \pb{F,\Phi_\alpha(\sigma)}
+\pb{F,\rho^\alpha(\sigma)}\Phi_\alpha(\sigma)
\right)
\approx \nonumber \\
&\approx&
\partial_\tau F+
\pb{F,H_0}+\int d\sigma
\rho^\alpha(\sigma) \pb{F,\Phi_\alpha(\sigma)}
 \ ,\nonumber \\
\end{eqnarray}
where $\approx$ means that this equation
holds on the constraint surface $\Phi_\alpha\approx 0$.
%%%%%%%%%%%%%%%%%%%%%%%%%%%%%%%%%%%%%%%%%
\section{Current algebra}\label{third}
In this section we will calculate
the current algebra of the principal
chiral model defined on the world-sheet
with general metric. Recall that these
currents are defined as
\begin{equation}
J_\alpha^A=E_M^A\partial_\alpha x^M \ . 
\end{equation}
In order to calculate the current
algebra we have to 
express these currents using
the canonical variables $x^M,p_M$. 
Namely we have to find
the dependence  $\partial_\tau
x^M$ on  $p_M,x^M$. Using
the Hamiltonian (\ref{mHT}) 
% will have
% important consequence for the
% existence of Lax connection. In fact,
% we know that the current $J_\alpha^A$
% is defined as
% \begin{equation}
% J_\alpha^A=E_M^A\partial_\alpha x^M \ . 
% \end{equation}
% However in the extended Hamiltonian 
% formalism $\partial_\tau x^M$ arrises from
% the first Hamiltonian equation of motion
we find the time derivative of $x^M$ as
\begin{eqnarray}
\partial_\tau x^M(\sigma)&=&
\pb{x^M(\sigma),H_T}=
\frac{(\lambda^++\lambda^-+
\rho^++\rho^-)}{2}
\frac{2\pi}{\sqrt{\lambda}}
G^{MN}p_N+\frac{\lambda^+-\lambda^-+
\rho^+-\rho^-}{2}
\partial_\sigma x^M=
\nonumber \\
&=&\left(\frac{\sqrt{-\gamma}}{\gamma_{\sigma\sigma}}
+\rho^0\right)\frac{2\pi}{\sqrt{\lambda}}
G^{MN}p_N+
\left(\frac{\gamma_{\tau\sigma}}{\gamma_{\sigma\sigma}}
+\rho^1\right)\partial_\sigma x^M \ 
\nonumber \\
\end{eqnarray}
and hence
% \begin{eqnarray}
% H=\int_0^{2\pi}d\sigma (\partial_\tau x^M
% p_M-\mL)=\nonumber \\ 
% =\int_0^{2\pi}
% d\sigma (-\frac{\sqrt{\lambda}}{4\pi}
% \sqrt{-\gamma}\gamma^{\tau\tau}
% E_M^A K_{AB} E^B_N\partial_\tau x^M\partial_\tau x^N
% +\sqrt{-\gamma}\gamma^{\sigma\sigma}
% E_M^A E_N^B K_{AB}\partial_\sigma x^M\partial_\sigma
% x^N)=\int_0^{2\pi}d\sigma \mH \ , 
% \nonumber \\
% \end{eqnarray}
% where 
% \begin{equation}
% \mH=-\frac{\sqrt{\lambda}}{4\pi}
% \sqrt{-\gamma}\gamma^{\tau\tau}
% J_\tau^AJ_\tau^BK_{AB}+
% \frac{\sqrt{\lambda}}{4\pi}
% \sqrt{-\gamma}\gamma^{\sigma\sigma}
% J_\sigma^A J_\sigma^B K_{AB} 
% \end{equation}
% We have expressed the Hamiltonian density
% in terms of the currents since we are interested
% in the Hamiltonian analysis of these currents. 
\begin{equation}\label{Jtau1}
J^A_\tau=
\frac{2\pi}{\sqrt{\lambda}}(-
\frac{1}{\sqrt{-\gamma}\gamma^{\tau\tau}}+\rho^0)
K^{AB}E_B^Mp_M+(-\frac{\gamma^{\tau\sigma}}
 {\gamma^{\tau\tau}}+\rho^1)E^A_M\partial_\sigma 
x^M \ , 
\end{equation}
where
\begin{equation}
E_B^ME_M^C=\delta_B^C \ , \quad 
E_M^BE_B^N=\delta_M^N \ . 
\end{equation}
Our  goal is  to calculate the algebra of currents
defined above. To do this we have to 
make presumptions about the dependence of the
Lagrange multipliers $\rho^\pm$ on the phase
space variables. Since we work in the theory where
the gauge symmetry is not fixed it is 
natural   to 
presume that the Lagrange multipliers
$\rho^\pm$ depend on $\sigma,\tau$ only
\cite{Govaerts:2002fq}:
$\rho^\pm=\rho(t,\sigma)$.
% The more
%general situation where $\rho^\pm$ 
%is a function of phase space variables 
%arrises in the case when we perform the gauge fixing
%of the constraints $T_0,T_1$. As we will dicsuss 
%below this fixing however also implies that we
%should rather calculate the Dirac brackets instead
%of Poisson brackets. We leave for future the discussion%
%of the gauge fixed form of the action and integrability.
Then it turns out that  it is useful to define
\begin{equation}
-\frac{1}{\sqrt{-\tgamma}
\tgamma^{\tau\tau}}=
-\frac{1}{\sqrt{-\gamma}
\gamma^{\tau\tau}}+\rho^0 \ ,  \quad 
-\frac{\tgamma^{\tau\sigma}}
{\tgamma^{\tau\tau}}=
-\frac{\gamma^{\tau\sigma}}
{\gamma^{\tau\tau}}+
\rho^1 \ 
\end{equation}
so that  we can write
\begin{equation}
J^A_\tau=-
\frac{2\pi}{\sqrt{\lambda}}
\frac{1}{\sqrt{-\tgamma}\tgamma^{\tau\tau}}
K^{AB}E_B^Mp_M-\frac{\tgamma^{\tau\sigma}}
 {\tgamma^{\tau\tau}}E^A_M\partial_\sigma 
x^M \  .  
\end{equation}
In the same way we also get 
\begin{equation}
\mH_T=
-\frac{1}{\sqrt{-\tgamma}
\tgamma^{\tau\tau}}
T_0-\frac{\tgamma^{\tau\sigma}}
{\tgamma^{\tau\tau}}T_1+
\rho^+_\lambda \pi_+^\lambda+
\rho^-_\lambda \pi_-^\lambda+
\rho^\xi \lambda_- \ . 
\end{equation}
Then the generalised action 
takes the form 
\begin{eqnarray}\label{Sgen}
S&=&\int d\sigma d\tau
[\partial_\tau x^M p_M+\partial_\tau
\lambda^+\pi_++\partial_\tau \lambda^-
\pi_-+\partial_\tau \xi \pi_\xi-
\mH_T]=\nonumber \\
&=&
-\frac{\sqrt{\lambda}}{4\pi}
\int d\sigma d\tau
\sqrt{-\tgamma}\tgamma^{\alpha\beta}
J_\alpha^A J_\beta^B K_{AB}+\nonumber \\
&+&\int d\sigma d\tau (
\partial_\tau\lambda^+
\pi_+^\lambda+
\partial_\tau\lambda^-\pi_-^\lambda+
\partial_\tau\lambda^\xi\pi_\xi-
\rho^+_\lambda \pi_+^\lambda-
\rho^-_\lambda \pi_-^\lambda
-\rho^\xi\pi_\xi) \ .  \nonumber \\
\end{eqnarray}
Then from (\ref{Sgen}) we
determine the  equations of motion for
current $J_\alpha$ in the form
\begin{equation}\label{eqJ}
\partial_\alpha [
\sqrt{-\tgamma}\tgamma^{\alpha\beta}
J_\beta^A]=0 \ . 
\end{equation}
Let us  now introduce the
Lax connection for given principal
model. We define its components as  
\begin{equation}\label{Lax}
\hJ_\alpha(\Lambda)=\frac{1}{1-\Lambda^2}
(J_\alpha-\Lambda\tgamma_{\alpha\beta}
\epsilon^{\beta\gamma}J_\gamma) \ , 
\end{equation}
where 
\begin{equation}
\epsilon^{\alpha\beta}=
\frac{e^{\alpha\beta}}{\sqrt{-\tgamma}} \ ,
\quad  
e^{\alpha\beta}=-e^{\beta\alpha} \ , \quad
e^{01}=1 \ 
\end{equation}
and where $\Lambda$ is a spectral parameter. 
Note that the Lax connection depends on $
\tgamma$ that is combination of the world-sheet
metric $\gamma$ and the gauge parameters $\rho^\pm$.
Then using the equations of motion (\ref{eqJ})
and also using the fact 
  that $J_\alpha^A$ 
obey the flatness condition
\begin{equation}\label{flatJ}
\partial_\alpha J^A_\beta-
 \partial_\beta  J^A_\alpha+
 J^B_\alpha J^C_\beta f_{BC}^A=0
 \end{equation}
 that follows from the definition of the
 currents $J_\alpha^A$ given in 
(\ref{defJ}) 
we can show that $\hJ$ is flat
\begin{equation}
\partial_\alpha \hJ^A_\beta-\partial_\beta
\hJ^A_\alpha+\hJ^B_\alpha \hJ^C_\beta
f_{BC}^A=0 \ . 
\end{equation}
% In fact, explicit calculation shows
% \begin{eqnarray}
% \partial_\tau \hJ^A_\sigma-\partial_\sigma
% \hJ^A_\tau=
% \frac{1}{1-\Lambda^2}
% (\partial_\tau J^A_\sigma-\partial_\sigma
% J^A_\tau)+
% \nonumber \\
% +\frac{\Lambda}
% {1-\Lambda^2}
% [-\partial_\tau[
% \sqrt{-\tgamma} \tgamma^{\tau\sigma}\tgamma^{\tau\alpha}
% J_\alpha^A-\partial_\sigma[\sqrt{-\tgamma}
% \tgamma^{\sigma\alpha}
% J_\alpha^A] \ , 
% \nonumber \\
% \hJ_\tau^B\hJ_\sigma^C f_{BC}^A=
% \frac{1}{(1-\Lambda^2)^2}
% (J_\tau^B+\Lambda\sqrt{-\tgamma}\tgamma^{\sigma\alpha}
% J_\alpha^B)(J_\sigma^C-
% \Lambda \sqrt{-\tgamma}\tgamma^{\tau\alpha}
% J_\gamma^C)f_{BC}^A=
% \nonumber \\
% \frac{1}{(1-\Lambda^2)^2}
% (J_\tau^BJ_\sigma^C-\Lambda^2\gamma
% (\gamma^{\sigma\sigma}\gamma^{\tau\tau}-
% (\gamma^{\tau\sigma})^2)
% J_\tau^BJ_\sigma^C)=\frac{1}{1-\Lambda^2}
% J_\tau^BJ_\sigma^Cf_{BC}^A
%  \nonumber \\
% \end{eqnarray}
% In other words the Lax connection
% obey the flatness condition 
% \begin{equation}
% \partial_\alpha \hJ^A_\beta
% -\partial_\beta \hJ^A_\alpha+
% \hJ_\alpha^B\hJ_\beta^C f_{BC}^A=0
% \end{equation}
% for general world-sheet metric.

Now we are going to 
 calculate the Poisson brackets
between currents $J_\alpha$.
Firstly, it is clear that
\begin{equation}
\pb{J^A_\sigma(\sigma),J^B_\sigma(\sigma')}=0 \ .
\end{equation}
On the other hand explicit calculation
gives
\begin{eqnarray}\label{Jtausigma}
& &\pb{J^A_\tau(\sigma),J^B_\sigma(\sigma')}=
% -\frac{2\pi}{\sqrt{\lambda}}
% \frac{1}{\sqrt{-\tgamma}\tgamma^{\tau\tau}(\sigma)}
% K^{AC}E_C^M(\sigma)\pb{p_M(\sigma),E_N^B(\sigma')
% \partial_{\sigma'}x^N(\sigma')}=
% \nonumber \\
% =\frac{2\pi}{\sqrt{\lambda}}
% \frac{1}{\sqrt{-\tgamma}\tgamma^{\tau\tau}(\sigma)}
% K^{AC}E_C^M(\sigma)\partial_M E^B(\sigma)
% \partial_\sigma X^N(\sigma)\delta(\sigma-\sigma')-
% \nonumber \\
% -\frac{2\pi}{\sqrt{\lambda}}
% \frac{1}{\sqrt{-\tgamma}\tgamma^{\tau\tau}(\sigma)}
% K^{AC}E_C^M(\sigma)E_M^B(\sigma')
% \partial_\sigma \delta(\sigma-\sigma')=
% \nonumber \\
\nonumber \\
&=&-\frac{2\pi}{\sqrt{\lambda}}
\frac{1}{\sqrt{-\tgamma}\tgamma^{\tau\tau}(\sigma)}
K^{AB}\partial_\sigma \delta(\sigma-\sigma')-
\frac{2\pi}{\sqrt{\lambda}}
\frac{1}{\sqrt{-\tgamma}\tgamma^{\tau\tau}(\sigma)}
K^{AC}E_C^M(\sigma)\partial_NE_M^B(\sigma)
\partial_\sigma x^N(\sigma) \delta(\sigma-\sigma')
+\nonumber \\
&+&\frac{2\pi}{\sqrt{\lambda}}
\frac{1}{\sqrt{-\tgamma}\tgamma^{\tau\tau}(\sigma)}
K^{AC}E_C^M(\sigma)\partial_M E^B_N(\sigma)
\partial_\sigma x^N(\sigma)\delta(\sigma-\sigma')=
\nonumber \\
% =-\frac{2\pi}{\sqrt{\lambda}}
% \frac{1}{\sqrt{-\tgamma}\tgamma^{\tau\tau}(\sigma)}
% K^{AB}\partial_\sigma \delta(\sigma-\sigma')+
% \frac{2\pi}{\sqrt{\lambda}}
% \frac{1}{\sqrt{-\tgamma}\tgamma^{\tau\tau}(\sigma)}
% K^{AC}E_C^M(\sigma) E_N^E E_M^D f_{ED}^B 
% \partial_\sigma x^N\delta(\sigma-\sigma')=
% \nonumber \\
&=&-\frac{2\pi}{\sqrt{\lambda}}
\frac{1}{\sqrt{-\tgamma}\tgamma^{\tau\tau}(\sigma)}
K^{AB}\partial_\sigma \delta(\sigma-\sigma')-
\frac{2\pi}{\sqrt{\lambda}}
\frac{1}{\sqrt{-\tgamma}\tgamma^{\tau\tau}(\sigma)}
J^C_\sigma f_{CD}^A K^{DB}\delta(\sigma-\sigma') \ , 
\nonumber \\
\end{eqnarray}
% where we have performed following
% maninuplation
% \begin{eqnarray}
% f(\sigma')\partial_\sigma\delta(\sigma-\sigma')=
% f(\sigma)\partial_\sigma\delta(\sigma-\sigma')+\nonumber \\
% +[\partial_\sigma f(\sigma)
% (\sigma'-\sigma)+\frac{1}{2}\partial^2_\sigma f
% (\sigma-\sigma')^2+
% \dots ]
% \partial_\sigma \delta(\sigma-\sigma')
% =f(\sigma)\partial_\sigma(\sigma-\sigma')+
% \partial_\sigma f(\sigma)\delta(\sigma-\sigma')
% \nonumber \\
% \end{eqnarray}
% using the fact that 
% that
% \begin{eqnarray}
% \int_0^{2\pi} d\sigma \partial^n_\sigma f(\sigma)
% (\sigma-\sigma')^n\partial_\sigma \delta(\sigma-\sigma')=
% \nonumber \\
% \partial^n_\sigma f(\sigma-\sigma')^n\delta(\sigma-\sigma')]_0^{2\pi}
% -\int_0^{2\pi}d\sigma 
% \partial_\sigma[\partial_\sigma^n f (\sigma-\sigma')^n]
% \delta(\sigma-\sigma')=
% \nonumber \\
% =-\int_0^{2\pi}d\sigma 
% \partial_\sigma^{(n+1)}f (\sigma-\sigma')^n
% \delta(\sigma-\sigma')-
% n\int_0^{2\pi}d\sigma 
% \partial_\sigma^{n}f (\sigma-\sigma')^{n-1}
% \delta(\sigma-\sigma')=0 \ , 
% \mathrm{for} \ n >1  \ , 
% \nonumber \\ 
% =-\partial_\sigma f(\sigma) \ , 
% \mathrm{for}  \ n=1 \  , \nonumber \\
% \end{eqnarray}
% where we have presumed that $\sigma'\neq 0,2\pi$. 
where we have used
\begin{equation}\label{UID}
f(\sigma')\partial_\sigma\delta(\sigma-\sigma')=
f(\sigma)\partial_\sigma \delta(\sigma-\sigma')
+\partial_\sigma f(\sigma)\delta(\sigma-\sigma')
\end{equation}
and also the  flatness
condition of the  current $J_\alpha^A$
(\ref{flatJ}) that implies 
\begin{eqnarray}
% \partial_\alpha J_\beta^A-\partial_\beta J_\alpha^A
% +J_\alpha^B J_\beta^C f_{BC}^A=0
% \Rightarrow 
% \nonumber \\
% =\partial_\alpha (E_M^A\partial_\beta x^M)-
% \partial_\beta (E_N^A\partial_\alpha x^N)
% +E^B_N E^C_M f_{BC}^A \partial_\alpha x^N\partial_
% \beta x^M=0 \Rightarrow 
% \nonumber \\
\partial_N E_M^A-\partial_M E^A_N+
E_N^B E_M^C f_{BC}^A=0 \ . 
\nonumber \\
\end{eqnarray}
For letter purposes we also determine the 
Poisson brackets between $J_\tau$ and $p_M,x^M$
\begin{eqnarray}
\pb{J_\tau^A(\sigma),p_M(\sigma')}
%-\frac{2\pi}{\sqrt{\lambda}}
%\frac{1}{\sqrt{-\tgamma}\tgamma^{\tau\tau}
%(\sigma)}K^{AB}\pb{E_B^N(\sigma),p_M(\sigma')}
%p_N(\sigma)-\nonumber \\
%-\frac{\tgamma^{\tau\sigma}(\sigma)}
%{\tgamma^{\tau\tau}(\sigma)}\pb{E_N^A(\sigma),
%p_M(\sigma')}\partial_\sigma x^N(\sigma)-
%\frac{\tgamma^{\tau\sigma}(\sigma)}
%{\tgamma^{\tau\tau}(\sigma)}
%E_M^A\partial_\sigma \delta(\sigma-\sigma')=
%\nonumber \\
&=&- \frac{2\pi}{\sqrt{\lambda}}
\frac{1}{\sqrt{-\tgamma}\tgamma^{\tau\tau}(\sigma)}
K^{AB}\partial_M E^N_B p_N(\sigma)-
\nonumber \\
&-&\frac{\tgamma^{\tau\sigma}(\sigma)}
{\tgamma^{\tau\tau}(\sigma)}
\partial_M E_N^A\partial_\sigma x^N
\delta(\sigma-\sigma') -E^A_M
\frac{\tgamma^{\tau\sigma}(\sigma)}
{\tgamma^{\tau\tau}(\sigma)}
\partial_\sigma \delta(\sigma-\sigma') \ , 
%=\partial_M J^A(\sigma) \delta(\sigma-\sigma')
\nonumber \\
\pb{J^A_\sigma(\sigma),p_M(\sigma')}&=&
\partial_M E^N_N\partial_\sigma x^N(\sigma)
\delta(\sigma-\sigma')+
E_M^A\partial_\sigma \delta(\sigma-\sigma') \  
\nonumber \\
%=
%\partial_M J_\sigma^A \delta(\sigma-\sigma')%
%\nonumber \\ 
\end{eqnarray}
and
\begin{equation}
\pb{J_\tau^A(\sigma),x^M(\sigma')}=
\frac{2\pi}{\sqrt{\lambda}}
\frac{1}{\sqrt{-\tgamma}\tgamma^{\tau\tau}(\sigma)}
K^{AB}E_B^M\delta(\sigma-\sigma') \ . 
\end{equation}
Then we obtain
\begin{eqnarray}\label{Jtautau}
& &\pb{J^A_\tau(\sigma),J^B_\tau(\sigma')}=
-\frac{2\pi}{\sqrt{\lambda}}
\pb{J^A_\tau(\sigma),E_D^M(\sigma')}
K^{BD}p_M(\sigma')\frac{1}{\sqrt{-\tgamma}
\tgamma^{\tau\tau}(\sigma')}-
\nonumber \\
&-& \pb{J_\tau^A(\sigma),J_\sigma^B(\sigma')}
\frac{\tgamma^{\tau\sigma}(\sigma')}
{\tgamma^{\tau\tau}(\sigma')}
-\frac{2\pi}{\sqrt{\lambda}}
\pb{J^A_\tau(\sigma),p_M(\sigma')}K^{BD}E_D^M(\sigma')
\frac{1}{\sqrt{-\tgamma}\tgamma^{\tau\tau}(\sigma')}=
\nonumber \\
&=&-\frac{2\pi}{\sqrt{\lambda}}
\frac{1}{\sqrt{-\tgamma}(\sigma)
\tgamma^{\tau\tau}(\sigma)}
J^C_\tau(\sigma) f_{CD}^A K^{DB}\delta(\sigma-\sigma')+
\nonumber \\
&+&
\frac{2\pi}{\sqrt{\lambda}}
\frac{\tgamma^{\tau\sigma}(\sigma)}
{\sqrt{-\tgamma}(\sigma)\tgamma^{\tau\tau}(\sigma)}
J^C_\sigma(\sigma) f_{CD}^A K^{DB}\delta(\sigma-\sigma')
+\nonumber \\
&+&\frac{2\pi}{\sqrt{\lambda}}
[\frac{1}{\sqrt{-\tgamma}(\sigma)\tgamma^{\tau\tau}(\sigma)}
\frac{\tgamma^{\sigma\tau}(\sigma')}
{\tgamma^{\tau\tau}(\sigma')}+
\frac{1}{\sqrt{-\tgamma}(\sigma')\tgamma^{\tau\tau}(\sigma')}
\frac{\tgamma^{\tau\sigma}(\sigma)}
{\tgamma^{\tau\tau}(\sigma)}]K^{AB}\partial_\sigma
\delta(\sigma-\sigma') \nonumber \\
\end{eqnarray}
using
\begin{eqnarray}
& & \partial_M E_A^N=-E_A^P
\partial_M E_P^BE_B^N \ , 
\nonumber \\
&  & E_C^M\partial_M E^N_D-E_D^M
\partial_M E^N_C=
%-E_C^M(\partial_M E_P^A -\partial_P E_M^A)
%E_A^N E_D^P=
%^\nonumber \\
%=E_C^M E_M^B E_P^F f_{BF}^A E_A^N E_D^P=
f_{CD}^AE_A^N \ . \nonumber \\
\end{eqnarray}
With the help of 
 the Poisson brackets 
(\ref{Jtausigma}) and (\ref{Jtautau})
we can determine the  Poisson 
bracket between the
spatial components of the Lax connection
for two different spectral parameters
$\Lambda,\Gamma$
\begin{eqnarray}
& &\pb{\hJ^A_\sigma(\sigma,\Lambda),
\hJ^B_\sigma(\sigma',\Gamma)}=
\nonumber \\
&=&-\frac{2\pi}{\sqrt{\lambda}}
\frac{K^{AB}(\Gamma+\Lambda)
}{(1-\Lambda^2)
(1-\Gamma^2)}
% +
% \Gamma\Lambda (\sqrt{-\tgamma}(\sigma)
% \tgamma^{\tau\sigma}(\sigma)
% +\sqrt{-\tgamma}(\sigma')\tgamma^{\tau\sigma}
% (\sigma'))+
% \nonumber \\
% +\Lambda\Gamma(-\sqrt{-\tgamma}
% (\sigma)\tgamma^{\tau\sigma}(\sigma)
% -\sqrt{-\tgamma}(\sigma')\tgamma^{\tau
% \sigma}(\sigma'))]
\partial_\sigma
\delta(\sigma-\sigma')-
\frac{2\pi}{\sqrt{\lambda}}
\frac{(\Lambda+\Gamma)}{(1-\Lambda^2)
(1-\Gamma^2)}J^C_\sigma(\sigma) f_{CD}^AK^{DB}
\delta(\sigma-\sigma')-\nonumber \\
&-&\frac{2\pi}{\sqrt{\lambda}}
\frac{\Lambda\Gamma}{(1-\Lambda^2)
(1-\Gamma^2)}(\sqrt{-\tgamma}\tgamma^{\tau\tau}
J_\tau^C+\sqrt{-\tgamma}\tgamma^{\tau\sigma}
J^C_\sigma )(\sigma)f_{CD}^AK^{DB}\delta(\sigma-\sigma')
 \ . \nonumber \\
\end{eqnarray}
The expression above  can be rewritten into
the form that on the right side
 contains the combinations of the spatial components
of the Lax connection  and term that is proportional
to the non-ultra-local
term $\partial_\sigma \delta(\sigma-\sigma')$
% Now we demand that the expression containing
% currents takes following form
% \begin{eqnarray}
% -\frac{2\pi}{\sqrt{\lambda}}
% (A\hJ^C_\sigma(\Lambda)-B\hJ^C_\sigma
% (\Gamma ) )f_{CD}^A K^{DB}\delta(\sigma-\sigma')
% \end{eqnarray}
% Using the explicit form of $\hJ$ and comparing with
% the expressions above we obtain set of two equations
% for unknown $A,B$
% \begin{eqnarray}
% \frac{A}{1-\Lambda^2}-\frac{B}{1-\Gamma^2}=\frac{\Gamma+
% \Lambda}{(1-\Gamma^2)(1-\Lambda^2)} \ , 
% \nonumber \\
% \frac{A \Lambda}{1-\Lambda^2}-\frac{B\Gamma}{1-\Gamma^2}
% =\frac{\Lambda \Gamma}{(1-\Gamma^2)(1-\Lambda^2)} \ ,
% \nonumber \\
% \end{eqnarray}
% that has solution
% \begin{equation}
% A=\frac{\Gamma^2}{(\Gamma^2-1)(\Lambda-\Gamma)} \ , 
% B=\frac{\Lambda^2}{(\Lambda^2-1)(\Lambda-\Gamma)} \ .
% \end{equation}
% In summary we obtain following Poisson bracket
\begin{eqnarray}\label{pbAL}
& &\frac{\sqrt{\lambda}}{2\pi}
\pb{\hJ^A_\sigma(\sigma,\Lambda),
\hJ^B_\sigma(\sigma',\Gamma)}=
-\frac{K^{AB}
(\Gamma+\Lambda)}{(1-\Lambda^2)
(1-\Gamma^2)}\partial_\sigma 
\delta(\sigma-\sigma')
-
\nonumber \\
&-&\frac{\Gamma^2}{(1-\Gamma^2)(\Lambda-\Gamma)} 
\hJ^C_\sigma(\Lambda)
f_{CD}^A K^{DB}
\delta(\sigma-\sigma')
-\frac{\Lambda^2}{(1-\Lambda^2 )(\Lambda-\Gamma )}
\hJ^C_\sigma (\Gamma )
f_{CD}^A K^{DB}\delta(\sigma-\sigma') \ . 
\nonumber \\
\end{eqnarray}
This result has precisely the same form
(if we use the tensor notation) as the
Poisson bracket given in 
\cite{Dorey:2006mx}. Since
 (\ref{pbAL}) does not explicitly depend on 
the world-sheet metric $\gamma_{\alpha\beta}$ and
the gauge parameters  $\rho^\pm$ we 
obtain that the Poisson bracket between
spatial components of the Lax connections 
is gauge invariant. This result has important
consequence for the integrability of the
theory. Recall that 
the monodromy matrix is defined
as \cite{Faddeev:1987ph,deVega:1983gy,Izergin:1980pe,
Maillet:1985ek,Maillet:1985ec}
\begin{equation}\label{Tdef}
\mT(l,\Lambda)=
P\exp \left(\int_l 
\hJ_\alpha(\Lambda) \frac{dx^\alpha}{d\xi}\right) \ , 
\end{equation}
where $l=(x^\alpha(\xi)\equiv (\tau'(\xi),
\sigma'(\xi)))$ is a curve on
 two dimensional cylinder, $\hJ_\alpha \frac{dx^\alpha}{d\xi}$
is an embedding of the Lax connection with 
spectral parameter $\Lambda$ on the curve
$l$  and $\xi$ is a
parameter that labels points on given curve. Finally
 $P$ in
(\ref{Tdef}) means  path ordering. We will
calculate the Poisson bracket of the monodromy
matrices that are defined at equal time 
$\tau$. In this case the curve
 $l$ is defined as 
\begin{eqnarray}\label{gammaf}
l=(\sigma'=\xi, \xi\in
(x_1,y_1)) \ . \nonumber \\
\end{eqnarray}
Then for the   curve (\ref{gammaf})
the monodromy matrix   $\mT$ takes the form
\begin{equation}\label{mTed}
\mT(\Lambda,x_1,y_1)=
P\exp\left(\int_{x_1}^{y_1}d\sigma'
\hJ_\sigma (\tau,\sigma' )\right) \ . 
\end{equation}
As was shown in  \cite{Faddeev:1987ph,deVega:1983gy,Izergin:1980pe,
Maillet:1985ek,Maillet:1985ec}
the Poisson brackets between elements
of the  monodromy
matrices $\mT(\Lambda,x_1,y_1)$ and
$\mT(\Gamma,x_2,y_2)$ are equal to
\begin{eqnarray}
\pb{\mT_{\alpha\beta}(x_1,y_1,\Lambda),
\mT_{\gamma\delta}(x_2,y_2,\Gamma)}
% \int d\sigma [(\frac{\delta \mT_{\alpha\beta}
% (\sigma_1,\sigma'_1,\Lambda)}
% {\delta x^B(\sigma)}
% \frac{\delta \mT_{\beta\gamma}
% (\sigma_2,\sigma'_2,\Gamma)}
% {\delta p_{B}(\sigma)}-\nonumber \\
% -\frac{\delta 
% \mT_{\alpha\beta}(\sigma_1,\sigma'_1,\Lambda)}
% {\delta p_{B}(\sigma)}
% \frac{\delta^L \mT_{\gamma\delta}
% (\sigma_2,\sigma'_2,\Gamma)}
% {\delta x^{B}(\sigma)})=\nonumber \\
=\int_{x_1}^{y_1}
d\sigma \int_{x_2}^{y_2}
d\sigma' \mT_{\alpha\omega_1}(x_1,\sigma,\Lambda)
\mT_{\omega_2\beta}(\sigma,y_1,\Lambda)
\times 
\nonumber \\
\times \pb{\hJ_{\omega_1\omega_2}(\sigma,\Lambda),
\hJ_{\rho_1\rho_2}(\sigma',\Gamma)}
\mT_{\gamma\rho_1}(x_2,\sigma',\Gamma)
\mT_{\rho_2\delta}(\sigma',y_2',\Gamma) \ ,
 \nonumber \\
\end{eqnarray}
where $\alpha,\beta,\dots$ label the matrix
indices of $\mT$. The form of the Poisson
bracket above shows a significance of the
Poisson bracket (\ref{pbAL}) for the
calculation of the Poisson brackets between
transition matrices and then between
the Poisson brackets of the conserved
 local charges. Recall that these local
 conserved charges arise from
 the expansion of the 
 trace of the monodromy
 matrix with respect to the corresponding
 spectral parameter
\cite{Dorey:2006mx}. Then since 
 (\ref{pbAL}) is gauge invariant
 and metric independent and takes 
 completely the same form as 
in \cite{Dorey:2006mx} we immediately
obtain that we can find an infinite number
of local conserved charges that are in involution
even on the string world-sheet with general
metric. 
  Of course all these
results depend on the presumption that
the Lagrange multipliers $\rho^\pm$
do not depend on the phase space variables.
In fact this is natural presumption  in
case when we do not fix the gauge since
the gauge symmetries imply the appearance 
of arbitrary functions in the  general
solutions of the equations of motion.
On the
other hand if we fix the gauge then
Lagrange multipliers can depend on the phase
space variables. Let us discuss this issue in 
more detail.  
%%%%%%%%%%%%%%%%%%%%%%%%%%%
%%%%%%%%%%%%%%%%%%%%%%%%%%%%%%%%%%%%%%%%%%%%%%%
% \section{Monodromy Matrix}
%%%%%%%%%%%%%%%%%%%%%%%%%%%%%%%%%%%%%%%%%%%%
 \subsection{Comments on the gauge fixing}
We  now present some comments 
considering the gauge fixing for the principal chiral
model defined on  
world-sheet with general metric. 
Recall that there are five
first class constraints defined above the equation
(\ref{mHT})
\begin{equation}
\Phi_\alpha\approx 0 \ .
\end{equation}
Fixing these constrains is achieved
by introducing  additional constraints
\cite{Henneaux:1992ig,Govaerts:1991gd,Govaerts:2002fq}
\begin{equation}
G_\alpha(\sigma)\approx 0 
\end{equation}
 that have nonzero Poisson bracket
with $\Phi_\alpha$ on the constraint surface
\begin{eqnarray}
\pb{G_\alpha(\sigma),\Phi_\beta(\sigma')}=
C_{\alpha\beta}(\sigma,\sigma') \ , \nonumber \\
% \pb{T_(\sigma),G_{+}(\sigma')}
% =C_{++}(\sigma\sigma') \ , 
% \pb{T_+(\sigma),G_{-}(\sigma')}=C_{+-} \ ,
% \nonumber \\ 
% \pb{T_-(\sigma),G_{+}(\sigma')}=
% C_{-+}(\sigma,\sigma') \ , 
% \pb{T_-(\sigma),G_{-}(\sigma')}=C_{--}
% (\sigma,\sigma') \ ,  \nonumber \\
\end{eqnarray}
where the matrix $C_{\alpha\beta}(\sigma,\sigma')$
is non-singular. Then the extended set
of constraints $\Phi_\alpha,G_\alpha$
consists the  second class constraints. 
In fact, the consistency of the theory demands
that these constraints are preserved
during the time-evolution of the system. Then
the Hamiltonian equation of motion for
$G_\alpha$ implies
\begin{eqnarray}
\frac{d}{d\tau}G_\alpha(\sigma)&=&\partial_\tau 
G_\alpha(\sigma)+
\pb{G_\alpha(\sigma),H_T}\approx\nonumber \\
% =\partial_\tau G_\alpha(\sigma)+
% \pb{G_\alpha(\sigma,H_0}+
% \int d\sigma'[\pb{G_\alpha(\sigma),\rho^\beta(\sigma')}
% \Phi_\beta(\sigma')+
% \rho^\beta(\sigma')\pb{G_\alpha(\sigma),
% \Phi_\beta(\sigma')}]\approx
% \nonumber \\
&\approx& 
\partial_\tau 
G_\alpha(\sigma)+\pb{G_\alpha(\sigma),
H_0}+
\int d\sigma'(
C_{\alpha\beta}
(\sigma,\sigma')\rho^\beta(\sigma'))=0 \ . 
\nonumber \\
\end{eqnarray}
Since $C_{\alpha\beta}$ is nonsingular this equation
determines $\rho^\alpha$ completely. In other
words the gauge symmetry is fixed.  
Note that after fixing the gauge the original
Hamiltonian is strongly zero.
 
As we have argued above all
 constraints form the collection
of the second class constraints.  
According to the general theory of 
constraint systems the proper way
how to deal with them is to introduce
the Dirac brackets. The Dirac
bracket between two phase
space functions $F,G$ takes the form
\begin{eqnarray}\label{DFG}
\pb{F(\sigma),
G(\sigma')}_D&=&
\pb{F(\sigma),G(\sigma')}
-\nonumber \\
&-&\int d\sigma''d\sigma'''\pb{F
(\sigma),\Phi_I(\sigma')}
C^{IJ}(\sigma'',\sigma''')\pb{\Phi_J(\sigma'''),
G(\sigma')} \ , 
\nonumber \\
\end{eqnarray}
where we have introduced following notation
\begin{eqnarray}
\Phi_1&=&T_+ \ , \quad
\Phi_2=T_- \ , \quad
\Phi_3=G_+ \ , \quad
\Phi_4=G_- \ , \nonumber \\
\Phi_5&=&\pi_+^\lambda \ , \quad 
\Phi_6=G_+^\lambda \ , \quad
\Phi_7=\pi_-^\lambda \ ,
\quad 
\Phi_8=G_-^\lambda \ , 
\nonumber \\
\Phi_9&=&\pi_\xi \ , \quad \Phi_{10}=
G_\xi \ , \quad  C_{IJ}(\sigma,\sigma')=
\pb{\Phi_I(\sigma),\Phi_J(\sigma')} \ , 
 \nonumber \\
\end{eqnarray}
and where in (\ref{DFG})
 the summation over $I,J$ 
 is understood. 
Note that $C^{IJ}$ is inverse of $C_{IJ}$ 
\begin{equation}
\int d\sigma''
C_{IJ}(\sigma,\sigma'')
C^{JK}(\sigma'',\sigma')=
\delta(\sigma-\sigma')\delta_I^K \ . 
\end{equation}
It can be shown the the Dirac brackets
of all constraints with any phase space
functions are zero. Consequently 
provided we use Dirac bracket instead of
the Poisson bracket, the second class
constraint $\Phi_I$ can be imposed even
before any calculations
\cite{Henneaux:1992ig,Govaerts:1991gd,Govaerts:2002fq}.
 We can then
explicitly solve these constraints $\Phi_I$
leading to the set of coordinates $z_m$
that parametrise reduced phase space 
that is equipped with the bracket structure
induced by Dirac bracket. 
Further, we should   find the generator
of the time evolution since 
when all the gauge freedom is fixed 
the  original
Hamiltonian strongly vanishes.

The advantage of the gauge fixing is that
we study the dynamics of the physical degrees
of freedom only. On the other hand
it is not completely clear whether 
the  gauge fixing condition
allows an existence of an infinite
number of conserved charges that
are in involution.
We postpone the  discussion of this problem
for future. 

% Generally the  gauge fixing implies that
% the  Lagrange multipliers $\rho^\alpha$ will
% depend  on the phase-space
% variables. 
%   As we have argued this leads
% to the complicated form of the time component
% of the current where we introduce $\partial_\tau
% x^M$ as follows from the Hamiltonian equation.
% However there is also third problem. If we
% completelly fix the gauge for covariant system
% we obtain that now the Hamiltonian is strongly
% zero. Then it follows that any evolution of
% the function with respect to $\tau$ is trivial.
% This is of course natural since $\tau$ is
% un-physical variable and the evolution with 
% respect to $\tau$ is merelly diffeomorphism
% transformation.  We come to this problem below.

On the other hand let us
consider  an example of the
partial gauge  fixing 
 that manifestly preserves
the integrability. To do
this we introduce following
gauge fixing condition
\begin{equation}\label{G+-}
G_+^\lambda=\lambda^+-f^+
 \ , \quad 
G_-^\lambda=\lambda^--f^- \ , \quad 
G_\xi=\xi-1 \ , 
\end{equation}
where $f^+,f^-$ generally depend on 
$\tau,\sigma$. Note that there is
still gauge freedom parametrised
by the Lagrange multipliers 
$\rho^\pm$.
 Then (\ref{G+-})
implies following non-zero matrix elements 
$C$
\begin{eqnarray}
C_{65}(\sigma,\sigma')&=&-C_{56}(\sigma',\sigma)=
\delta(\sigma-\sigma') \ , \nonumber \\ 
C_{87}(\sigma,\sigma')&=&-C_{87}(\sigma',\sigma)
=\delta(\sigma-\sigma') \ , 
\nonumber \\  
C_{9,10}(\sigma,\sigma')&=&
-C_{10,9}(\sigma',\sigma)=\delta(\sigma-\sigma') \ .
\nonumber \\
\end{eqnarray}
The requirement that the
gauge fixing constraints (\ref{G+-})
have to be preserved during the
time evolution determines following
values of the Lagrange multipliers
$\rho_\pm^\lambda,\rho_\xi$:   
\begin{equation}
\rho_\pm^\lambda=0 \ , 
\quad \rho_\xi=0 \ .   
\end{equation}
Then  
 the partially fixed 
 Hamiltonian takes the form 
\begin{eqnarray}
H&=&H_0+\int d\sigma
(\rho^+\Phi_++\rho^-\Phi_-)=
\nonumber \\
&=&\int d\sigma 
\left[(f^++\rho^+)T_++
(f^-+\rho^-)T_-\right] \ . 
\nonumber \\
\end{eqnarray}
Let us now discuss the
 consequence of the
gauge fixing (\ref{G+-})
on the form of  the current
algebra. Since we have
not fixed $\rho^\pm$
we can again presume that 
$\rho^\pm$ depend on 
$\tau,\sigma$ only. 
According to the general
analysis we should calculate
the Dirac bracket of the
current algebra. However
it is easy to see from 
(\ref{G+-}) that the 
Dirac bracket of the canonical
variables $p_M,x^N$ coincide
with the Poisson bracket. Then
the current algebra takes precisely
the same form as was determined
in the previous section when
we replace $\gamma_{\alpha\beta}$
with corresponding combinations of
$f^\pm$ defined in (\ref{G+-}).
It is also clear that the Poisson
bracket of the Lax connection
takes the same form as in
(\ref{pbAL}). According to the
discussion given in the previous
section this result ensures an 
integrability of the theory. 
In summary principal chiral model
with fixed world-sheet metric is
integrable. 

It is also important to
stress that we have not fixed the
gauge symmetry generated by  $T_\pm$.
Then the classical analysis should proceed
along the standard treatment of the
constraint systems. Explicitly, 
the constraints $T_\pm$ are imposed on the
system after performing all calculations. 
We also  demand that the physical
observables have vanishing Poisson brackets
with the generators of the gauge
transformations on the constraint surface. 
The equivalent requirement on the physical
observables is that they commute with the
BRST generator corresponding to given gauge
symmetries. In fact we have shown in our previous
paper \cite{Kluson:2007gp}
 that  the monodromy matrix 
is BRST invariant. This
result also implies
an existence of an 
infinite number of BRST invariant conserved
charges.

\section{Nambu-Gotto form of principal model}\label{fourth}
In this section we  consider the Nambu-Gotto
 form of the principal model. It is clear that this
form of the string action arises from the Polyakov
form of the action through the integration of the
world-sheet metric. On the other hand it is not
completely clear whether this procedure does
not spoil integrability of the theory. For that reason
we mean that it is useful to 
study the integrability of this model as well. 

The Nambu-Gotto principal chiral
model action takes the form
\begin{equation}\label{SNG}
S=\frac{\sqrt{\lambda}}{2\pi}
\int d\sigma d\tau \mL \ ,  \quad 
\mL=-
\sqrt{-\det \bA} \ , 
% -\frac{1}{2}
% \varepsilon^{\alpha\beta}B_{\alpha\beta}
\end{equation}
where
\begin{equation}
% B_{\alpha\beta}=B_{AB}J^A_\alpha J^B_\beta \ , 
% B_{AB}=-B_{BA} \ , 
\bA_{\alpha\beta}=
K_{AB}J^A_\alpha J^B_\beta \ ,  \quad 
J_\alpha=g^{-1}\partial_\alpha g=
J^A_\alpha T_A  \ . 
\end{equation}
Let us now consider the variation of
the group element $g$ in the form
\[\delta g=g^{-1}\delta X,\
\delta X=\delta X^A T_A \]
that implies following variation of 
the current $J$ 
\begin{equation}
\delta J_\alpha=-\delta X J_\alpha+
J_\alpha \delta X+
\partial_\alpha \delta X  \ , 
\end{equation}
or explicitly
\begin{equation}\label{deltaJA}
\delta J^A_\alpha
=J^B_\alpha f_{BC}^A\delta X^C+
\partial_\alpha \delta X^A \ . 
\end{equation}
Then the variation of the action
(\ref{SNG}) under the variation
(\ref{deltaJA}) takes the 
form
\begin{eqnarray}
\frac{2\pi}{\sqrt{\lambda}}
\delta S_{NG}&=&
\int d\sigma d\tau
\frac{1}{2}\delta \bA_{\alpha\beta}\bAi^{\beta\alpha}
\sqrt{-\det\bA}=
% \delta J^A_\alpha K_{AB}J^B_\beta
% \bAi^{\beta\alpha}\sqrt{-\det\bA}= 
\nonumber \\
&=&-\delta X^C
\left(\partial_\alpha[ K_{CB}J^B_\beta 
\bAi^{\beta\alpha}\sqrt{-\det\bA}]-\right.
\nonumber \\
&-& \left. f_{CB}^A J^B_\alpha K_{AD}J^D_\beta
\bAi^{\beta\alpha}\sqrt{-\det\bA}
\right) \ . 
\nonumber \\
\end{eqnarray}
The last term in the expression
above vanishes 
%due to the fact that
%\begin{equation}
%J^B_\alpha f_{BC}^A K_{AD}J^D_\beta\bA^{\beta\alpha}
%=-J^B_\alpha f_{BD}^AJ^D_\beta
%K_{AC} \bAi^{\beta\alpha}=0
%\end{equation}
due to the antisymmetry of $f_{AB}^C$ and
symmetry of $\bAi^{\alpha\beta}$. If we now demand
that the variation of the action vanishes
for on-shell configuration 
we  obtain the equation of motion
for $J^A$ in the form
\begin{equation}\label{eqNG}
\partial_\alpha[ J^A_\beta 
\bAi^{\beta\alpha}\sqrt{-\det\bA}]=0 \ . 
\end{equation}
Let us now consider following
current
\begin{eqnarray}\label{LaxNG}
\hJ_\sigma(\Lambda)&=&\frac{1}{1-\Lambda^2}
[J_\sigma^A+\Lambda \bAi^{\tau\alpha}J_\alpha^A
\sqrt{-\det\bA}] \ ,  \nonumber \\
\hJ_\tau(\Lambda)&=&\frac{1}{
1-\Lambda^2}[J_\tau^A-\Lambda \bAi^{\sigma\alpha}J_\alpha^A
\sqrt{-\det\bA}] \ . \nonumber \\ 
\end{eqnarray}
The components of the current given
above obey the equation
\begin{eqnarray}\label{Lax1}
& &\partial_\tau \hJ_\sigma^A-
\partial_\sigma \hJ_\tau^A=
\frac{1}{1-\Lambda^2}
[\partial_\tau J^A_\sigma-\partial_\sigma J_\tau^A 
+\nonumber \\
&+&\Lambda \left(\partial_\tau[\bAi^{\tau\alpha}J_\alpha^A
\sqrt{-\det\bA}]+
\partial_\sigma [\bAi^{\sigma\alpha}
J_\alpha^A\sqrt{-\det\bA}]\right)]=
\nonumber \\
&=&
-\frac{1}{1-\Lambda^2}
J^B_\tau J^C_\sigma f_{BC}^A \ , 
\nonumber \\
\end{eqnarray}
where we have used the equation of 
motion (\ref{eqNG}) and the flatness of the
current $J_\alpha^A$. On the other hand
let us also calculate
\begin{eqnarray}\label{Lax2}
\hJ^B_\tau \hJ^C_\sigma f_{BC}^A&=&
\frac{1}{(1-\Lambda^2)^2}
[J^B_\tau J^C_\sigma f_{BC}^A+
\Lambda J_\tau^B J^C_\sigma f_{BC}^A 
\bAi^{\tau\sigma}\sqrt{-\det\bA}-
\nonumber \\
&-&\Lambda \bAi^{\sigma\tau}
J_\tau^B J_\sigma^C f_{BC}^A
\sqrt{-\det\bA}-
\Lambda^2 \bAi^{\sigma\alpha}J_\alpha^B
\bAi^{\tau\beta}J_\beta^C f_{BC}^A\sqrt{-\det\bA}]=
\nonumber \\
&=&\frac{1}{(1-\Lambda^2)^2}
(1-\Lambda^2)J^B_\tau J^C_\sigma f_{BC}^A
\nonumber \\
\end{eqnarray}
using the fact that
\begin{eqnarray}
&-&\Lambda^2\bAi^{\sigma\alpha}J_\alpha^B
\bAi^{\tau\beta}J_\beta^C f_{BC}^A(\sqrt{-\det\bA})^2=
%=\Lambda^2(\bAi^{10}J_0^B+\bAi^{11}J_1^B)
%(\bAi^{00}J_0^C+\bAi^{01}J_1^C)f_{BC}^A
%(\sqrt{-\det\bA})^2=
%\nonumber \\
%-(\bAi^{10}\bAi^{01}J^B_0J^C_1
%+\bAi^{11}\bAi^{00}J_1^BJ_0^C)f_{BC}^A
%\sqrt{-\det\bA}=
% \nonumber \\
%&=&-\frac{\Lambda^2}{\det\bA}J^B_\tau J^C_\sigma f_{BC}^A
%\det\bA=
-\Lambda^2 J^B_\tau J^C_\sigma f_{BC}^A \ .
\nonumber \\
\end{eqnarray}
If we collect (\ref{Lax1}) with
(\ref{Lax2}) we  obtain
an important result
\begin{equation}\label{Laxeqf}
\partial_\alpha \hJ_\beta^A-
\partial_\beta \hJ_\alpha^A+
\hJ^B_\alpha \hJ^C_\beta f_{BC}^A=0 \ . 
\end{equation}
In other words the components
of the current (\ref{LaxNG}) form the
Lax connection and (\ref{Laxeqf}) implies
an integrability of the theory. 
%Finally we have to return to the term
%in the Lagrangian that contains $B$.
%\begin{eqnarray}
%\delta\frac{1}{2}
%(B_{\alpha\beta}\epsilon^{\alpha\beta})=
%\delta J_\alpha^A B_{AB} J^B_\beta \epsilon^{\alpha\beta}=
%\nonumber \\
%-\delta X^A\partial_\alpha[\epsilon^{\alpha\beta}
%B_{AB}J^B_\beta]-\delta X^A J^C_\alpha f_{AC}^B
%B_{BD}J^D_\beta=
%\nonumber \\
%=-\delta X^A B_{AB}
%(\partial_0 J^B_1-\partial_1 J^B_0
%J^C_0 J^D_1 f_{CD}^B) 
%\nonumber \\
%\end{eqnarray}
%where we have used the fact that ussualy $B$ is
%proportional to $K$. In any case this expression does
%not contribute to the equation of motion. On the other
%hand it is clear that it could generally affect the
%Hamiltonian analysis. 
%Let us start the Hamiltonian analysis in case of zero
%$B$ field. 

Now we proceed to the calculations
of the Poisson brackets of the currents.
As usual we have to develop the Hamiltonian
formalism for the model 
(\ref{SNG}).  
% Let us again write
% the current as
% \begin{equation}
% J^A_\alpha=E^A_M\partial_\alpha x^M
% \end{equation}
As follows from the form of the action (\ref{SNG})
 the momentum conjugate to $x^M$ takes the form
\begin{eqnarray}\label{pmNG}
p_M=\frac{\delta S}{\delta\partial_\tau x^M}=
-\frac{\sqrt{\lambda}}{2\pi}G_{MN}\partial_\alpha
x^N \bAi^{\alpha \tau}\sqrt{-\det\bA} \ .
\nonumber \\
\end{eqnarray}
Then it is easy to determine the primary constraint
of the theory 
\begin{eqnarray}
% p_M G^{MN}p_N=
% -(\frac{\sqrt{\lambda}}{2\pi})^2
% \bAi^{\tau\alpha}\partial_\alpha x^M G_{MN}
% \bAi^{\tau\beta}\partial_\beta x^N \det \bA=
% \nonumber \\
% =-\bAi^{\tau\tau}\det \bA=-\bA_{\sigma\sigma}
% \Rightarrow 
\Phi_0=\frac{\pi}{\sqrt{\lambda}}
p_M G^{MN}p_N+\frac{\sqrt{\lambda}}{4\pi}
\partial_\sigma x^M G_{MN}\partial_\sigma x^N=0 \ . 
\nonumber \\
\end{eqnarray}
At the same time  
 (\ref{pmNG}) implies 
the second primary constraint 
\begin{equation}
\Phi_1=p_M\partial_\sigma x^M=0 \ . 
% -\frac{\sqrt{\lambda}}
% {2\pi}
% (\bA_{\sigma \sigma}\bAi^{\sigma\tau}+
% \bA_{\sigma\tau}\bAi^{\tau\tau})\sqrt{-\det\bA}=0
\end{equation}
Since these constraints have exactly the same
form as the constraints found in section
(\ref{second}) it is clear that their time
evolutions do not imply any secondary constraints.
Moreover, the original 
 Hamiltonian density takes the form
\begin{eqnarray}
\mH_0=p_M\partial_\tau x^M-\mL=0 
\end{eqnarray}
as we can expect from the fact that the
Nambu-Gotto action is diffeomorphism invariant.
Then according to the general principles of the
dynamics of the constraint systems
\cite{Henneaux:1992ig,Govaerts:1991gd,Govaerts:2002fq}
the total Hamiltonian density takes the form
\begin{equation}
\mH_T=\lambda^0 \Phi_0+
\lambda^1\Phi_1 \ . 
\end{equation}
Then it follows that $\partial_\tau x^M$
is equal to
\begin{equation}
\partial_\tau x^M=
\pb{x^M,H_T}=
\lambda^0 \frac{2\pi}{\sqrt{\lambda}}
G^{MN}p_N+\lambda^1 \partial_\sigma x^M \ .
\end{equation}
Using this result we express 
the current $J_\tau^A$ as a function of
the canonical variables $p_M,x^M$
\begin{equation}
J^A_\tau=E^A_M\partial_\tau x^M=
\lambda^0\frac{2\pi}{\sqrt{\lambda}}
K^{AB}E_B^Np_N+\lambda^1 E^A_M\partial_\sigma x^M
\ . 
\end{equation}
This form of the current 
is the same as the time component
of the current given in
(\ref{Jtau1}) if we
perform an identification
\begin{equation}
\lambda^0=-\frac{1}{\sqrt{\tgamma}
\tgamma^{\tau\tau}} \ , \quad  
\lambda^1=-\frac{\tgamma^{\tau\sigma}}
{\tgamma^{\tau\tau}} \ .  
\end{equation}
Without lost of generality we can
presume that the  Lagrange multipliers
$\lambda^{0,1}$
 do not depend on the phase space
variables and we obtain the Poisson
brackets of the current in the form
\begin{eqnarray}
\pb{J^A_\sigma(\sigma),J^B_\sigma(\sigma')}&=&0 \ ,
\nonumber \\
 \pb{J^A_\tau(\sigma),J^B_\sigma(\sigma')}&=&\lambda^0\frac{
2\pi}{\sqrt{\lambda}}[
K^{AB}\partial_\sigma \delta(\sigma-\sigma')+
J^C_\sigma f_{CD}^A K^{DB}\delta(\sigma-\sigma')] \ , 
\nonumber \\
\pb{J^A_\tau(\sigma),J^B_\tau(\sigma')}&=&
\lambda_0\frac{2\pi}{\sqrt{\lambda}}
J^C_\tau f_{CD}^A K^{DB}\delta(\sigma-\sigma')-
\lambda^1\frac{2\pi}{\sqrt{\lambda}}
J^C_\sigma f_{CD}^A K^{DB}\delta(\sigma-\sigma')
+\nonumber \\
&+&\frac{2\pi}{\sqrt{\lambda}}
(\lambda^0(\sigma)\lambda^1(\sigma')+
\lambda^0(\sigma')\lambda^1(\sigma))
K^{AB}\partial_\sigma
\delta(\sigma-\sigma') \ .  \nonumber \\
\end{eqnarray}
Let us now proceed to the calculation
of the Poisson bracket of the spatial
component of the Lax connection. 
Using  (\ref{pmNG})
we can write it in the form
\begin{eqnarray}
\hJ_\sigma=\frac{1}{1-\Lambda^2}
[J_\sigma^A-\frac{2\pi}{\sqrt{\lambda}}
\Gamma K^{AB}E_B^M p_M] \ .
\nonumber \\
\end{eqnarray}
% using the fact that
% \begin{equation}
% -\frac{2\pi}{\sqrt{\lambda}}
% K^{AB}E_B^Mp_M=\bAi^{\tau\alpha}J^B_\alpha 
% \sqrt{-\det\bA} \ ,
% \end{equation}
Then with the help of the
formula
\begin{equation}
\pb{J^A_\sigma(\sigma),p_M(\sigma')}=
E^A_M(\sigma)
\partial_\sigma\delta(\sigma-\sigma')+
\partial_\sigma x^N(\sigma)
\partial_M E_N^A(\sigma)\delta(\sigma-\sigma')
\end{equation}
we  obtain
\begin{eqnarray}
& &\pb{\hJ^A_\sigma
(\Lambda)(\sigma),\hJ^B_\sigma
(\Gamma)(\sigma')}
% \frac{1}{(1-\Lambda^2)(1-\Gamma^2)}
% (-\frac{2\pi}{\sqrt{\lambda}}
% \Gamma K^{BC}E_C^M(\sigma')\pb{J_\sigma^A(\sigma),p_M(\sigma')}
% +\nonumber \\
% &+&\frac{2\pi}{\sqrt{\lambda}}
% \Lambda K^{AD}E_D^N(\sigma)\pb{J^B_\sigma(\sigma'),p_N(\sigma)}+
% (\frac{2\pi}{\sqrt{\lambda}}
% )^2\Lambda \Gamma K^{AC}K^{BD}
% \pb{E_C^M(\sigma)p_M(\sigma),E_D^N(\sigma')p_N(\sigma')} )=
% \nonumber \\ 
% =\frac{1}{(1-\Gamma^2)
% (1-\Lambda^2)}[-\Gamma\frac{2\pi}{\sqrt{\lambda}}K^{BA}\partial_\sigma
% \delta(\sigma-\sigma')+\frac{2\pi}{\sqrt{\lambda}}
% \Gamma  K^{BC}E_C^M(\sigma)[\partial_N E_M^A-
%  \partial_M E_N^A]
% \partial_\sigma x^N(\sigma)\delta(\sigma-\sigma')+
% \nonumber \\
% \Lambda\frac{2\pi}{\sqrt{\lambda}}
% K^{AD}E_D^N(\sigma)[E_N^B(\sigma')\partial_{\sigma'}
% \delta(\sigma'-\sigma)+\partial_N 
% E^B_M(\sigma')\partial_{\sigma'}x^M(\sigma')
% \delta(\sigma-\sigma')]-
% \nonumber \\
% -(\frac{2\pi}{\sqrt{\lambda}})^2
% \Lambda \Gamma K^{AC}K^{BD}[-E^M_C\partial_M E^N_D+
% E_D^M\partial_M E^N_C]p_N\delta(\sigma-\sigma')]=
% \nonumber \\
% =-\frac{\Gamma+\Lambda}{(1-\Gamma^2)(1-\Lambda^2)}
% \frac{2\pi}{\sqrt{\lambda}}
% K^{AB}\partial_\sigma \delta(\sigma-\sigma')
% -\frac{2\pi}{\sqrt{\lambda}}\Gamma K^{BC}E_C^M
% E_N^E E_M^F\partial_\sigma x^N f_{EF}^A\delta(\sigma-\sigma')
% +\nonumber \\
% +\Lambda \frac{2\pi}{\sqrt{\lambda}}
% K^{AD}E_D^M E_N^E E^F_M f_{EF}^B\partial_\sigma
% x^N\delta(\sigma-\sigma')+
% \nonumber \\
% +\frac{\Lambda\Gamma}{(1-\Gamma^2)(1-\Lambda^2)}
% (\frac{2\pi}{\sqrt{\lambda}})^2
% K^{AC}K^{BD}f_{CD}^G E_G^N p_N\delta(\sigma-\sigma')=
% \nonumber \\
=-\frac{\Gamma+\Lambda}{(1-\Gamma^2)(1-\Lambda^2)}
\frac{2\pi}{\sqrt{\lambda}}
K^{AB}\partial_\sigma \delta(\sigma-\sigma')-\nonumber \\
&-&\frac{2\pi}{\sqrt{\lambda}}
\frac{\Gamma+\Lambda}{(1-\Gamma^2)(1-\Lambda^2)}
 J^C_\sigma f_{CD}^A K^{DB}\delta(\sigma-\sigma')
-\frac{2\pi}{\sqrt{\lambda}}
\frac{\Lambda\Gamma}{(1-\Lambda^2)(1-\Gamma^2)}
J^C_\tau f_{CD}^AK^{DB}\delta(\sigma-\sigma')
\nonumber \\
\end{eqnarray}
that again can be  rewritten
in the more natural form 
\begin{eqnarray}
& &\frac{\sqrt{\lambda}}{2\pi}
\pb{\hJ^A_\sigma(\sigma,\Lambda),
\hJ^B_\sigma(\sigma',\Gamma)}=
-\frac{K^{AB}
(\Gamma+\Lambda)}{(1-\Lambda^2)
(1-\Gamma^2)}\partial_\sigma 
\delta(\sigma-\sigma')
-
\nonumber \\
&-&\frac{\Gamma^2}{(1-\Gamma^2)(\Lambda-\Gamma)} 
\hJ^C_\sigma(\Lambda)
f_{CD}^A K^{DB}
\delta(\sigma-\sigma')
-\frac{\Lambda^2}{(1-\Lambda^2 )(\Lambda-\Gamma )}
\hJ^C_\sigma (\Gamma )
f_{CD}^A K^{DB}\delta(\sigma-\sigma') \ . 
\nonumber \\
\end{eqnarray}
Following the careful analysis given in 
\cite{Dorey:2006mx} we can now argue that
the action (\ref{SNG}) is integrable
and after appropriate regularisation it
possesses an infinite number of conserved
charges that are in involution. 
%%%%%%%%%%%%%%%%%%%%%%%%%%%%%%%%%%%%%%%%%%%%%%%%
\section{Conclusion}\label{fifth}
This short note was devoted to the calculation
of the algebra of the current for principal chiral
model coupled to two dimensional gravity. We have
determined the algebra of currents that
depend on the world-sheet metric and on the
gauge parameters. Then we have calculated 
the Poisson bracket of the spatial component
of the Lax connection and we have shown
that this result does not depend on the metric
and gauge parameters. This fact implies an
existence of infinite number of gauge
independent conserved charges that are in involution.
We have also shown that the  same conclusions
are valid for the 
Nambu-Gotto form of the action. 

The extension of this work is as follows. 
We mean that it is very interesting problem to
 study  the integrability of the bosonic
string on $AdS_5\times S^5$ in the 
uniform light-cone gauge
\cite{Arutyunov:2006gs,Arutyunov:2005hd}.
 Even if it was shown in 
\cite{Arutyunov:2004yx}  that it is possible
to define the Lax connection for the
gauge fixed form of the action as well, the
calculation of the Poisson bracket of the
spatial component of the Lax connection  has
not been performed  yet.
% One can expect that
%this calculation will be very complicated
%as was demonstrated in  \cite{Das:2005hp}
%in case of different gauge fixing. 
We think that it would be very useful to 
know whether the gauge fixed form of the
action possesses infinite number of the
integrals of motions that are in involution.
We hope to return to this problem in future.

\section*{Acknowledgements}

I would like to thank J. Govaerts for
very interesting and stimulating correspondence.
This work  was supported in part by the Czech Ministry of
Education under Contract No. MSM
0021622409, by INFN, by the MIUR-COFIN
contract 2003-023852 and , by the EU
contracts MRTN-CT-2004-503369 and
MRTN-CT-2004-512194, by the INTAS
contract 03-516346 and by the NATO
grant PST.CLG.978785.

\end{document}